\documentclass[preprint,12pt]{elsarticle}

\usepackage{amssymb}
\usepackage{amsthm}
\usepackage{amsmath}
\usepackage{amsmath}
\usepackage{verbatim}

\PassOptionsToPackage{hyphens}{url}
\usepackage{hyperref}
\usepackage{xurl}
\usepackage[utf8]{inputenc}
\usepackage[T1]{fontenc}

\journal{Nuclear Instruments and Methods in Physics Research - section A}

\begin{document}

\begin{frontmatter}



\title{Temperature Dependence of the Time Resolution in a SiPM-Readout Plastic Scintillator for Cosmic--Ray Applications}


\author[inst1]{José Reyes Castillo}
\author[inst1]{Saúl Aguilar Salazar}
\author[inst1]{Diego Mauricio Gomez Coral\corref{cor1}}
\ead{dgomezco@fisica.unam.mx}
\cortext[cor1]{Corresponding author}

\affiliation[inst1]{organization={Instituto de Física, Universidad Nacional Autónoma de México},
            addressline={Circuito de la investigación científica, Ciudad Universitaria}, 
            city={CDMX},
            postcode={04500}, 
            state={CDMX},
            country={Mexico}}


\begin{abstract}
Balloon- and space-borne cosmic--ray experiments employ plastic scintillators read out by silicon photomultipliers (SiPMs) to achieve picosecond--level time resolutions for triggering and particle identification. The performance of these systems can be affected by temperature variations encountered in flight. In this work, a time--of--flight (TOF) prototype consisting of a BC418 plastic scintillator bar coupled to Onsemi MICROFC-30050 SiPMs was constructed and tested under a controlled thermal environment between $\text{–20}$ and 20\,°C. Electrons from a $^{90}\text{Sr}$ source were used as a beam, and a dedicated differential preamplifier and coincidence triggering were implemented to study the detector response. A minimum time resolution of 160\,ps was achieved at an overvoltage of 3\,V, remaining stable across the tested temperature range and uniform along the scintillator bar.

\end{abstract}




\end{frontmatter}


\section{Introduction}
\label{sec:intro}

\subsection{Motivation}

The direct detection of cosmic rays has entered a precision measurement era, with experiments in space such as AMS-02\,\cite{AMS_TOF1997}, CALET\,\cite{CALET2013}, and DAMPE\,\cite{DAMPE2017} on the International Space Station reporting uncertainties on the level of a few percent. Their time--of--flight (TOF) or Plastic Scintillator Detector (PSD) systems, equipped with photomultiplier tubes (PMTs), are used for triggering and particle identification, achieving time resolutions of hundreds of picoseconds (ps) in AMS-02. Currently, new balloon--borne experiments are being deployed, such as GAPS\,\cite{GAPS} for antinuclei detection and HELIX\,\cite{HELIX} for light--isotope identification, which include Silicon Photomultipliers (SiPMs) in their TOF systems, improving time resolution. Moreover, forthcoming experiments such as HERD\,\cite{HERD2014} will extend detection capabilities for high-energy cosmic rays near the knee, with improved timing layers and the use of plastic fibers and SiPMs for particle tracking. CubeSat projects are also testing lightweight scintillator--SiPM assemblies as cost--effective payloads for high-altitude orbits\,\cite{ComSAD2021}. In addition, planned next--generation experiments such as ALADInO\,\cite{ALADInO} and AMS-100\,\cite{AMS100} have more ambitious goals, aiming to be located at Lagrange Point L2, and to achieve time resolution on the order of a few tens of picoseconds. These experiments attempt to extend measurements of cosmic--ray fluxes and composition with unprecedented precision, requiring detectors capable of operating reliably in the harsh conditions of near space. Compact, low--power, and efficient systems of plastic scintillators coupled to SiPMs are therefore of significant interest in this field and require a detailed study.

Improved timing capabilities are essential for accurate reconstruction of particle velocity and for identification of particle species through TOF techniques. It is known that fast plastic scintillators are the best active material for TOF detectors due to their short decay times and high light yields, two crucial parameters affecting time resolution\,\cite{Leo1994}. When coupled to SiPMs, these detectors offer significant advantages in time resolution over traditional PMTs, thanks to their fast response, high gain, and high efficiency. They also show substantial benefits for space and balloon missions, including low operating voltages, immunity to magnetic fields, mechanical robustness, and rapidly decreasing costs\,\cite{Gundacker_2020}. 

\subsection{Silicon Photomultipliers (SiPMs)}

A silicon photomultiplier (SiPM) is a solid--state device composed of hundreds or thousands of Geiger--mode avalanche photo--diodes (micro--cells), which produce a fast signal when one or multiple photons hit it\,\cite{DRenker_2009}.  The breakdown voltage ($\text{V}_{\text{bd}}$) corresponds to the reverse--bias voltage ($\text{V}_{\text{bias}}$) at which Geiger-mode avalanche multiplication begins in the micro--cells and depends on the internal structure\,\cite{Gundacker_2020, ACERBI201916}. The overvoltage ($\text{V}_{\text{ov}}$) is the excess bias voltage above the breakdown voltage, which determines the strength of the electric field in the avalanche region ($\text{V}_{\text{ov}}=\text{V}_{bias}-\text{V}_{\text{bd}}$). It governs the main performance parameters of the device, including gain ($G$), photon detection efficiency (PDE), and correlated noise. Increasing the overvoltage raises the gain, increases trigger avalanche probability, resulting in a higher PDE and an increased noise signal. The gain is defined as the amount of charge flowing per avalanche and is expressed as $G=C_{cell}\text{V}_{\text{ov}}/e=C_{cell}(\text{V}_{bias}-\text{V}_{\text{bd}})/e$, where $C_{cell}$ is the effective capacitance of a micro-cell. The total output signal is the sum of the responses from all fired micro--cells, yielding a gain comparable to PMTs ($10^5-10^6$). The gain stability depends directly on maintaining a constant overvoltage. Thus, temperature--induced changes in breakdown voltage must be compensated to keep $G$ constant. Due to their inherent solid-state nature, SiPMs are affected by thermal conditions. In particular, an average current due to thermally induced avalanches, called dark current, is present. The dark count rate (DCR) quantifies the corresponding number of individual dark pulses detected per second.

An important characteristic of SiPMs for TOF detectors is the fast internal avalanche amplification, which provides excellent timing information for the photon's arrival at the photodetector. In this work, we present an experimental study of the time resolution of a plastic scintillator detector prototype read out by SiPMs, focusing on its performance under different thermal conditions. In Section\,\ref{sec:setup}, the experimental setup is introduced, and in Section\,\ref{sec:results}, the measurement results and analysis are presented. Conclusions are summarized in Section\,\ref{sec:conclusions}.

\section{Experimental setup}
\label{sec:setup}

Two types of experimental setups were used in this work. On the one hand, a setup was designed to study the variation of the SiPM's breakdown voltage with temperature and the time resolution of the TOF counter as a function of temperature (Setup\,1). On the other hand, the time resolution as a function of the source position in the TOF counter at room temperature was investigated with a slightly different arrangement (Setup\,2). In both cases, four Onsemi MICROFC-30050-SMT\,\cite{ONSMicroC} SiPMs, soldered to a PCB in an array, were used as photodetectors. An external high-precision power supply, Keithley 2280S-60-3\,\cite{Tektronix2280S}, provided the bias voltage for the four SiPMs. 

\begin{figure*}[t]
\begin{center}
\begin{tabular}{ll}
\includegraphics[width=0.7\linewidth]{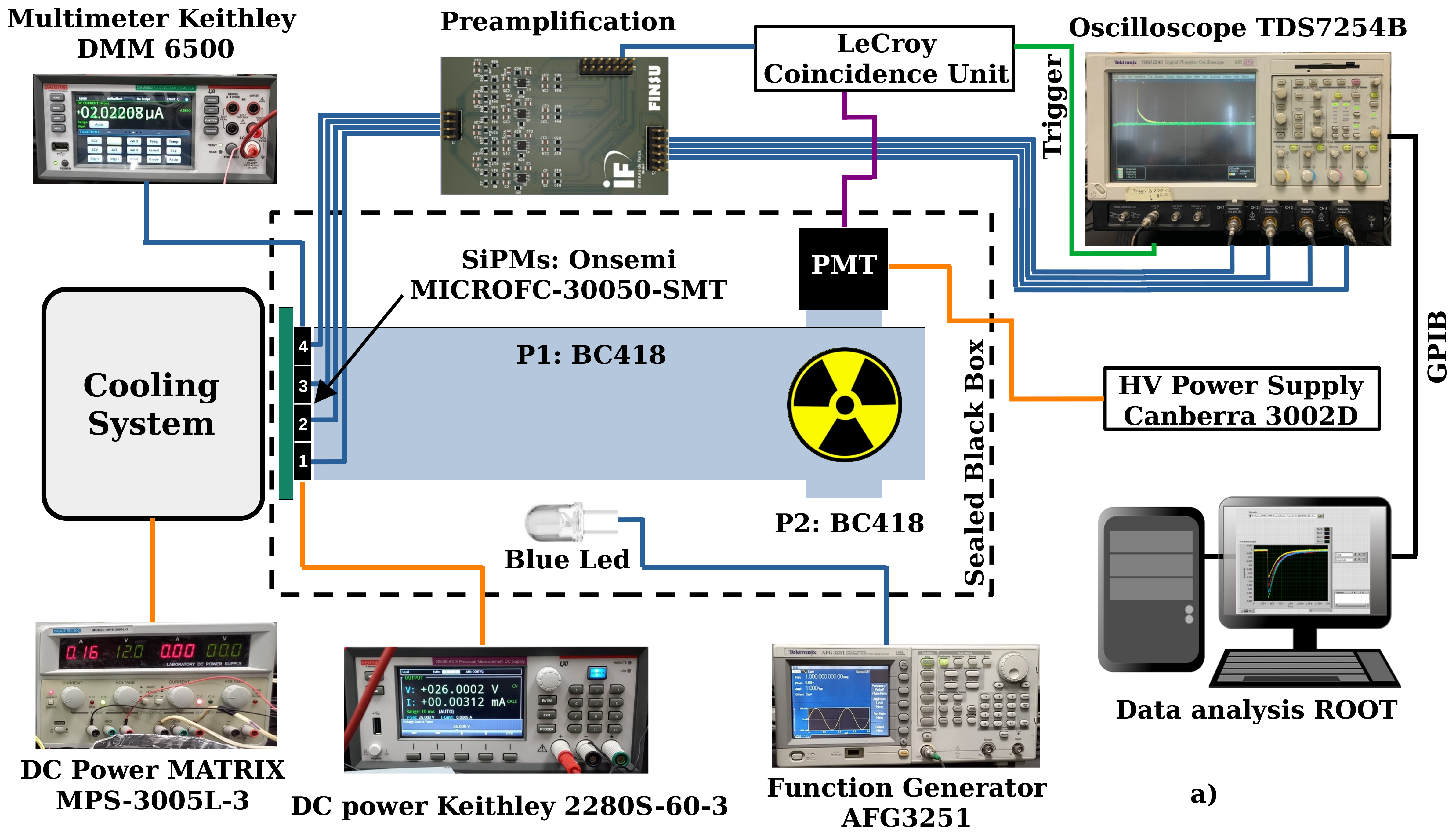}
& \includegraphics[width=0.25\linewidth]{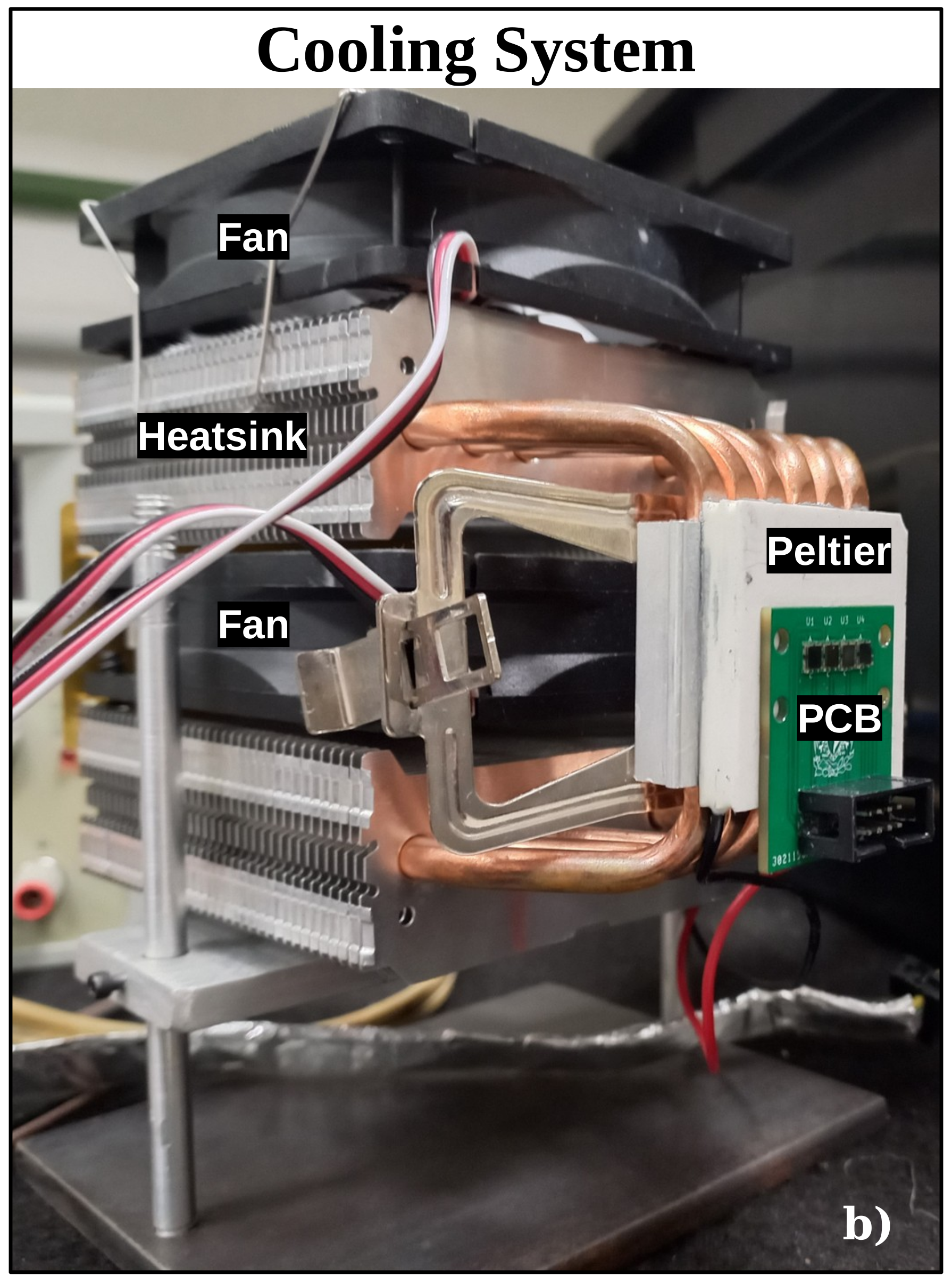}
\end{tabular}
\caption{(a) Diagram of the different experimental setups used in this work. (b) Photograph of the cooling system used in Setup\,1. See text for details.}
\label{fig1}
\end{center}
\end{figure*}

A fast--timing plastic scintillator bar BC-418\,\cite{BC418}, labeled P1, with dimensions 100 mm in length, 16 mm in width, and 3 mm in thickness, was tested. The bar was optically coupled to the array of four SiPMs with optical grease and wrapped with Tyvek\,{\footnotesize\textregistered} and black tape. A second small plastic scintillator bar, labeled P2, of the same reference as P1, was coupled to a Photomultiplier Hamamatsu R14755U-100\,\cite{HamamatsuR14755U100}, and positioned just below P1. The signal from the PMT in coincidence with a SiPM's signal acted as a trigger, as shown in Figure\,\ref{fig1}\,(a). An electron source ($^{90}\text{Sr}$) with a collimator of 1\,mm diameter was put above P1 and P2 and used as a beam to test the counter (see Figure\,\ref{fig1}\,(a)). Electrons from $^{90}\text{Sr}$ and its daughter $^{90}\text{Y}$ have a continuous energy spectrum up to 2.2 MeV, and can be considered MIPs in plastic scintillator above 1\,MeV\,\cite{Arfaoui:2013995}.

\subsection{Setup\,1: Temperature Dependence}\label{subsec:setup1}

The cooling system is integrated by a Peltier module TEC2-19006 12V 6A\,\cite{TEC219006}, coupled to a CPU cooler as shown in Figure\,\ref{fig1}\,(b). The Peltier module and the cooler are fed by a High DC Power Supply, MATRIX 3005L-3\,\cite{MPS300xL}, with a maximum current of 5\,A and a voltage of 30\,V. By varying the supply current, temperatures down to -30 degrees Celsius can be achieved at the contact point, and the system can maintain a stable temperature for several hours, enabling steady data collection. The temperature was continuously monitored by three K--type thermocouples connected to digital thermometers located near the Peltier module and the PCB, with a precision of $\pm$\,0.3\,°C. Measurements were performed only after verifying that thermal equilibrium had been achieved.

The characteristic I--V curve was obtained at each temperature to measure the breakdown voltage as a function of temperature. A blue LED (470 nm) was pointed at a SiPM using a mechanical holder with a collimator of 3\,mm diameter, and its intensity was modulated by a function generator, Tektronix--AFG3251\,\cite{TektronixAFG3000}. The detection efficiency expected for the LED wavelength is 25 to 35\%, for 2.5 to 5\,V overvoltage, with a reduction of only 5\% from the maximum efficiency reported by the vendor\,\cite{ONSMicroC}. The PCB+LED system was housed in a sealed black box to prevent any light background. The bias voltage was manually varied, and a digital multimeter, Keithley DMM6500\,\cite{KeithleyDMM6500} measured the current (see Figure\,\ref{fig1}\,(a)). 

For time resolution as a function of temperature, a TOF counter consisting of the SiPMs--PCB plus P1 was used along with the cooling system described above. By maintaining the $^{90}\text{Sr}$ source in a fixed point and recording the waveforms from SiPMs after the trigger, it will be possible to measure time differences between SiPM pairs at each temperature. 

\subsection{Setup\,2: Position Dependence}\label{subsec:setup2}

No active cooling system was used to measure the time resolution as a function of the source position along the plastic scintillator bar P1. In this case, two identical SiPM--PCBs are coupled to P1 on opposite sides. Time differences between opposite SiPM pairs were measured at a room temperature of 20\,°C using the same trigger system as in Setup\,1 while varying the source position along the longitudinal axis.  

\subsection{Preamplification}

\begin{figure}[t]
\begin{center}
\includegraphics[width=7.0cm]{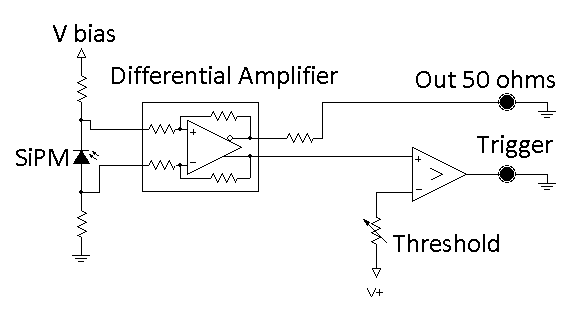}
\caption{Diagram of the differential preamplification used in this work.}
\label{fig2}
\end{center}
\end{figure}

A dedicated differential preamplification was developed to read out SiPM signals independently. A circuit diagram for one channel is shown in Figure\,\ref{fig2}. Both the SiPM bias voltage and the pulse signal are sent through a twisted--pair cable to be received by the differential amplifier. Once decoupled from DC, the signal is amplified 3 times\,\footnote{Although the circuit was built to amplified 10 times, impedance coupling reduced this amplification to 3 times.} by the AD8351\,\cite{AD8351}, which has a differential output, one output is used to be read by the digitizer, and the other is connected to an ultra-fast comparator MAX912\,\cite{MAX912}. The comparator's output was used as a trigger when it exceeded a reference voltage threshold.

\subsection{Data Acquisition System}
The DAQ system for this experiment consists of an Oscilloscope TDS7254B\,\cite{TektronixCSA7404B} with a bandwidth of 2.5 GHz and a total sampling of 20 GS/s divided into four channels at 5 GS/s per channel. The oscilloscope was connected to the computer via a GPIB interface, and a LabVIEW\,\cite{LabVIEW} program was written to read and save all waveforms in a TDMS file format. 

\section{Data analysis and results}
\label{sec:results}

\subsection{Breakdown voltage vs temperature}
\label{subsec:bdvtemp}

At lower temperatures, phonon scattering in the semiconductor decreases, allowing for higher charge carrier mobility and increasing the probability of a breakdown avalanche to occur at lower bias voltage\,\cite{Gundacker_2020}. Empirically, a linear behavior is expected between $\text{V}_{\text{bd}}$ and temperature (T), with a slope that depends on the SiPM design and materials\,\cite{ONSMicroC}. Therefore, it is important to measure this dependence for the SiPMs used in this work.

\begin{figure*}[!hbt]
\begin{center}
\begin{tabular}{ll}
\includegraphics[width=0.45\linewidth]{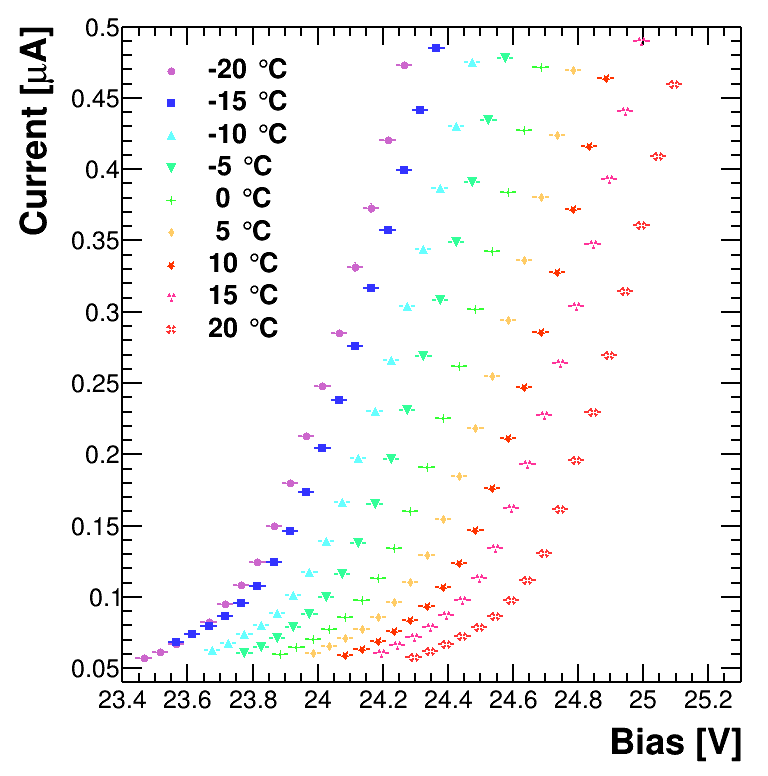}
& \includegraphics[width=0.45\linewidth]{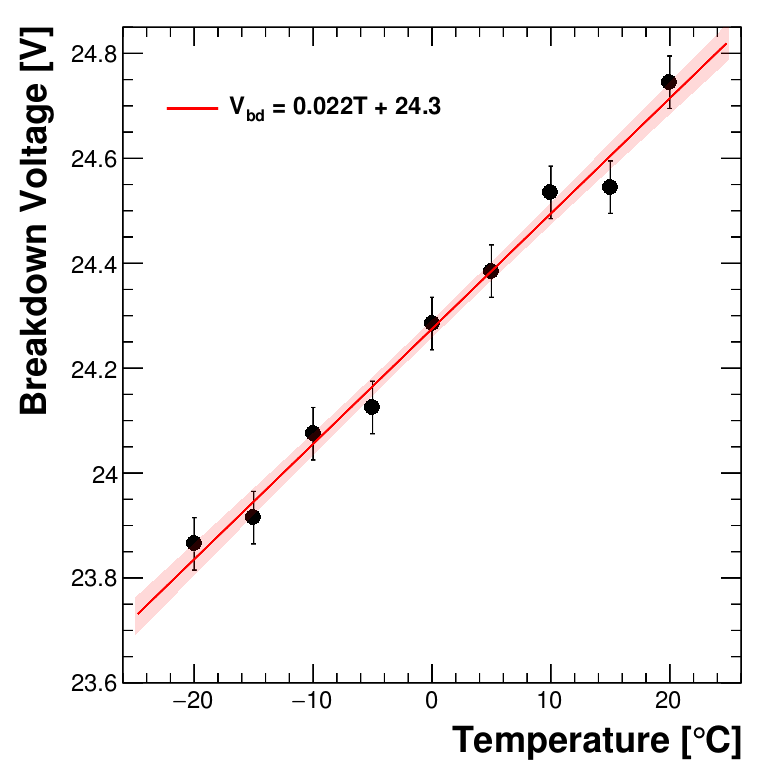}
\end{tabular}
\caption{The left plot shows measured current vs reverse-bias voltage at each temperature for Onsemi MICROFC-30050-SMT SiPMs. The right plot illustrates the linear relationship between breakdown voltage and temperature for the same SiPM.}
\label{fig3}
\end{center}
\end{figure*}

Following Setup\,1 described in Sec.\,\ref{subsec:setup1}, a blue LED with a fixed frequency and amplitude was pointed directly at the SiPM using a collimator. The current as a function of the bias voltage was measured as shown on the left panel of Figure\,\ref{fig3}. It can be seen how the current starts to increase dramatically after a specific breakdown voltage. These measurements were repeated at nine temperatures from -20 to 20\,°C. To obtain the particular point of $\text{V}_{\text{bd}}$, a similar process as in\,\cite{GOYAL20223750, OTTE2017106} was followed, where the centroid of the (dI/dV)/I vs $\text{V}_{bias}$ distribution was taken. After obtaining all $\text{V}_{\text{bd}}$ values, a clear linear behavior was observed. An example is shown on the right panel of Figure\,\ref{fig3}, with the resulting linear fit parameters. As can be observed, the calculated slope, i.e., the temperature coefficient, is 22.0\,$\pm$\,1\,mV/°C, very close to the value of 21.5\,mV/°C reported by the vendor\,\cite{ONSMicroC}. Furthermore, the intercept parameter from the linear fit is 24.3\,V, just about the minimum value 24.2\,V from\,\cite{ONSMicroC}. The red shadow band in Figure\,\ref{fig3} (right) is the 68\% confidence interval. The breakdown voltage was also obtained by an alternative independent numerical method, using a bilinear fit to log(I) vs $\text{V}_{bias}$ and finding the intersection (knee)\,\cite{NAGY201755}. A deviation close to 1\% was observed in the temperature coefficient between the two methods. Therefore, a 1\% systematic uncertainty to the temperature coefficient of the breakdown voltage was assigned. Similar values were found for all SiPMs used in this experiment. These measurements are crucial for determining the SiPMs' overvoltage at various temperatures. 


\subsection{Wave signal analysis}\label{subsec:wave}

\begin{figure*}[t]
\begin{center}
\begin{tabular}{ll}
\includegraphics[width=0.44\linewidth]{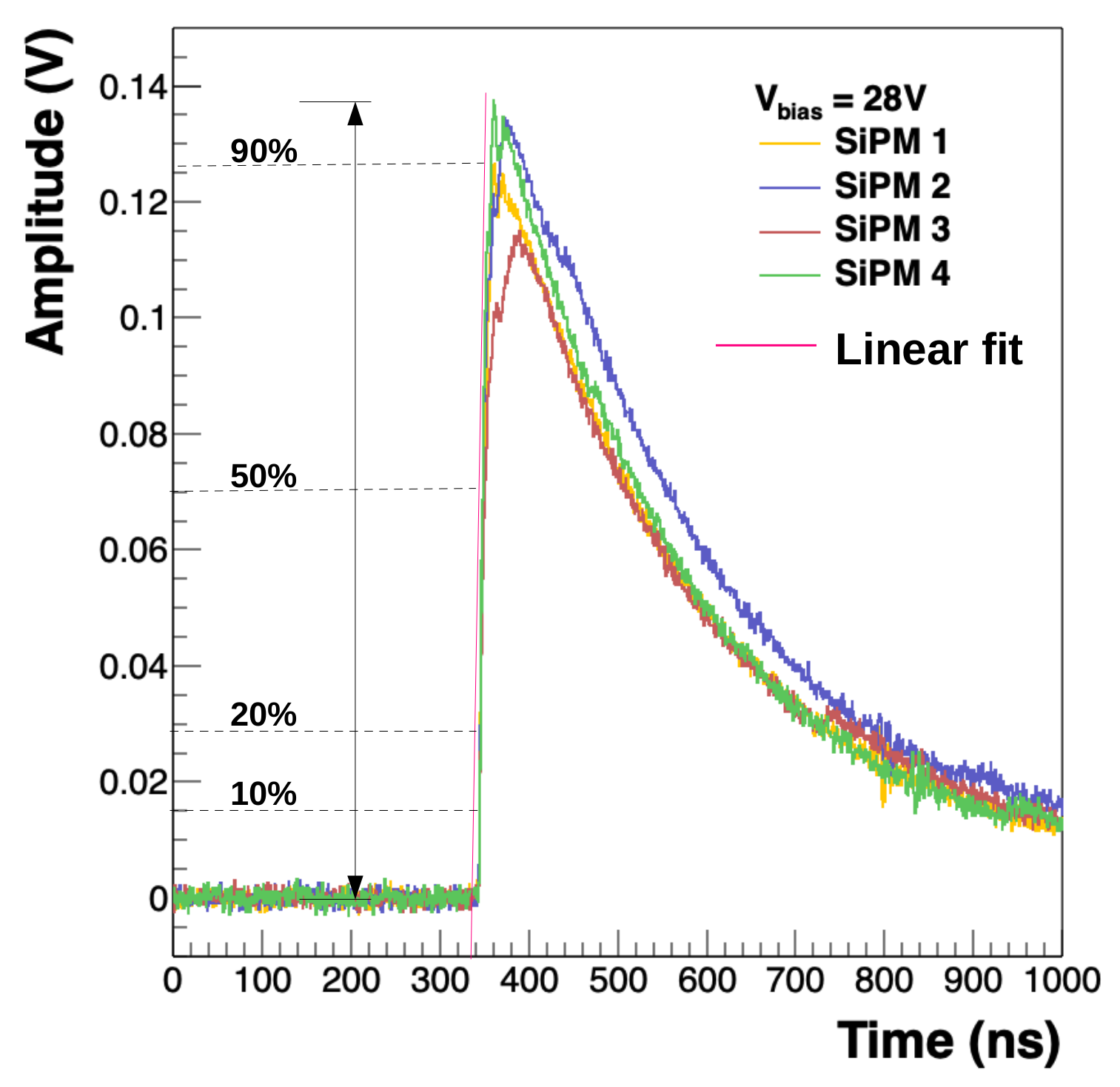}
& \includegraphics[width=0.46\linewidth]{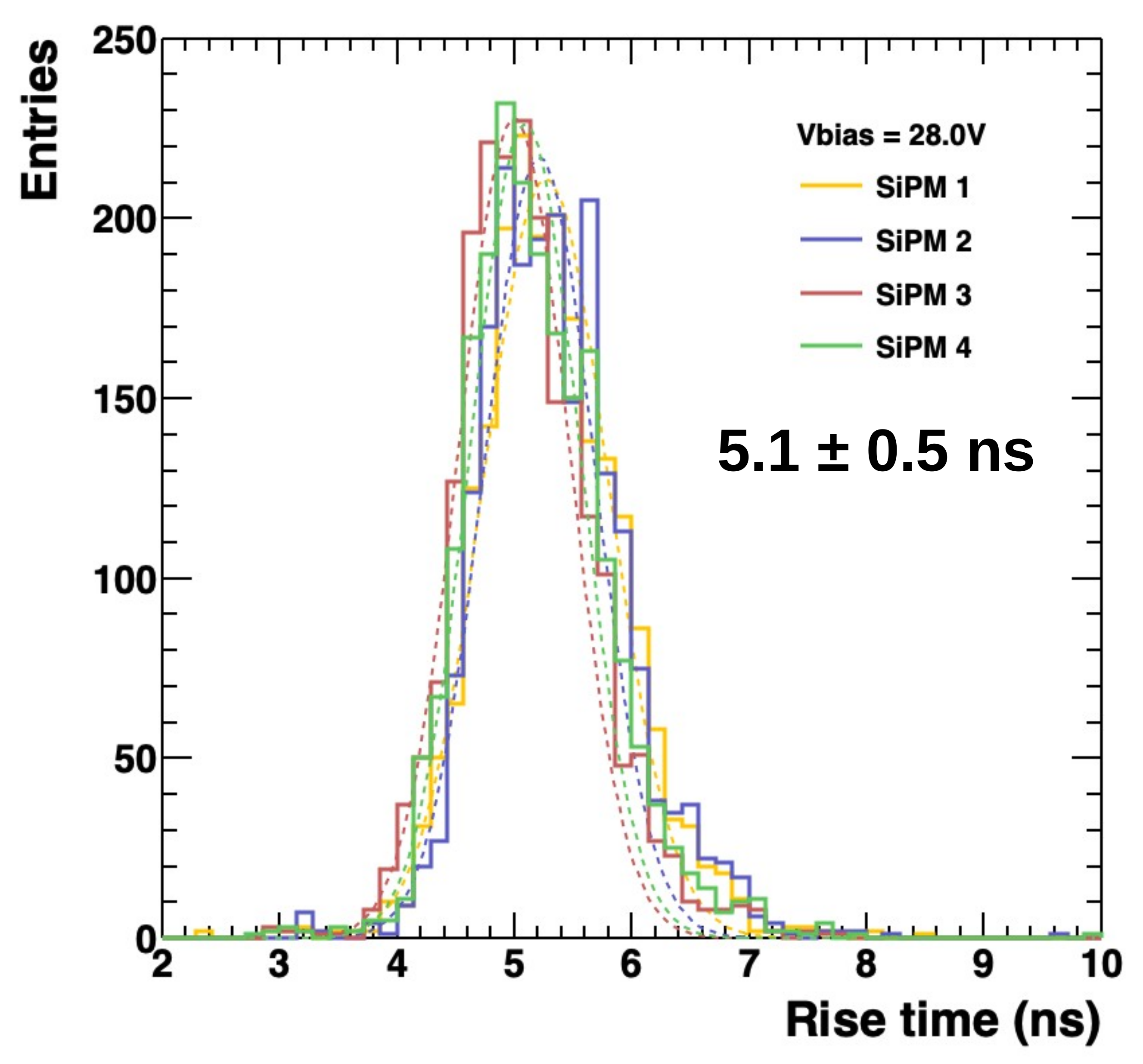}
\end{tabular}
\caption{On the left, representative waveforms of every SiPM signal with the linear fit to the leading edge and threshold levels at 10\%, 20\%, 50\%, and 90\% are shown. On the right, the measured rise time distributions are shown with their corresponding Gaussian fits.}
\label{fig4}
\end{center}
\end{figure*}

Time information from TOF counters in Setup\,1 and\,2 was obtained from the SiPM voltage signal in time, i.e., the waveform. After all waveforms from a run were saved in a TDMS file format, a free interface\,\cite{libtdms}, which was adapted for the specific needs of this analysis, was used to read these TDMS files and fill every waveform in a ROOT histogram (see Figure\,\ref{fig4}\,left). Processing waveforms as histograms enabled standardized extraction of pulse-shape parameters. 

For the determination of timing information, a software implementation of the constant fraction discrimination (CFD) technique was employed\,\cite{GEDCKE1967377}. A linear fit was applied to the leading edge of each pulse, and time stamps were extracted at specific amplitude fractions. The left panel of Figure\,\ref{fig4} illustrates a representative waveform together with the linear fit and the threshold levels at 10\%, 20\%, and 50\% of the pulse maximum. Time stamps were evaluated at these fractions, and it was observed that the 10\% and 20\% levels yielded comparable precision. The 20\% fraction was therefore adopted as the reference level for subsequent time--difference measurements.

The right panel of Figure\,\ref{fig4} shows the rise time distributions extracted from the CDF analysis of each SiPM signal, at 28\,V bias and 20\,°C. The rise time is the time for the pulse to rise from 10\% to 90\% of its maximum amplitude. From Gaussian fits to all distributions, consistent results were obtained, yielding an average rise time of 5\,ns.  

\subsection{Time resolution}
\label{subsec:timeres}

Time resolution is measured from the time difference distribution between two SiPM timestamps. If $t_1$ and $t_2$ represent these timestamps, then the time difference is $\Delta t = (t_1-t_2)$. From the spread in $\Delta t$ distribution labeled $\sigma_{\Delta t}$, different definitions of time resolution are obtained. For a two-detector coincidence system, the coincidence time resolution (CTR)\,\cite{instruments6010014, PANDA_TOF_TDR} is 

\begin{equation}
\sigma_{CTR} = \sigma_{\Delta t}  = \sqrt{\sigma_{t_1}^2 + \sigma_{t_2}^2}.
\end{equation}

Therefore, when both detectors have equal uncorrelated timing uncertainties, the individual time resolution\,\cite{Alici_2018} can be expressed as 

\begin{equation}
\sigma_{ITR} = \frac{\sigma_{\Delta t}}{\sqrt{2}}.
\end{equation}

The average time of both detectors is $t = (t_1+t_2)/2$. For TOF systems, it is usual to define the time resolution as the resolution of the average time\,\cite{instruments6010014, PANDA_TOF_TDR}, i.e.,

\begin{equation}
\sigma_{t} = \frac{\sqrt{\sigma_{t_1}^2 + \sigma_{t_2}^2}}{2} = \frac{\sigma_{\Delta t}}{2}.
\end{equation}

For measurements made using Setup\,1, where only one SiPM-PCB is coupled to one side of the plastic scintillation bar, $t_1$ and $t_2$ are time measurements from different SiPM pairs in the same array. In the case of Setup\,2, two SiPM-PCBs are coupled to opposite sides of P1, therefore $t_1$ and $t_2$ correspond to time measurements from different and opposite SiPMs.

\begin{figure}[t]
\begin{center}
\includegraphics[width=7.0cm]{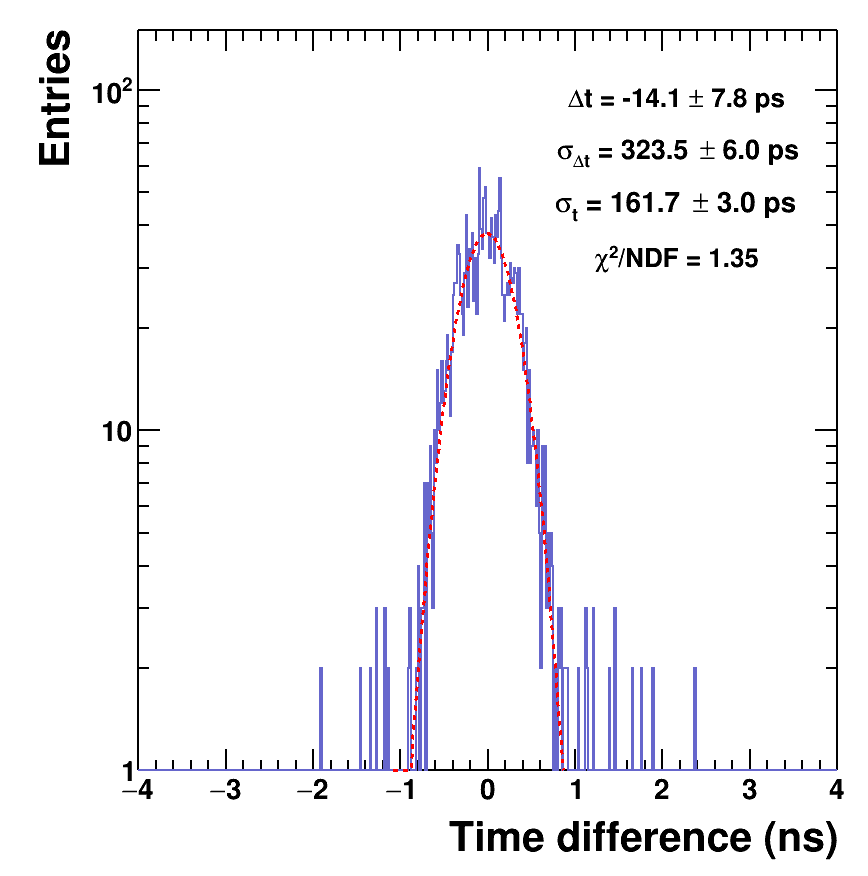}
\caption{Time difference distribution between two SiPMs measured with Setup\,2, and using a CFD at 20\%.} 
\label{fig5}
\end{center}
\end{figure}

Figure\,\ref{fig5} shows the $\Delta t$ distribution measured with Setup\,2, maintaining SiPMs at 3\,V overvoltage and a room temperature of 20\,°C. This distribution can be fitted by a Gaussian function as shown by the red line in Figure\,\ref{fig5}, with a good $\chi^2$/NDF=1.37. The mean value obtained from the fit is $\text{-14.1}\pm\text{7.8}$\,ps, reflecting a small systematic offset between SiPM timestamps, and a standard deviation (CTR) of $\text{323.5}\pm\text{6.0}$\,ps. For this example, the time resolution of the average results in a value of $\sigma_{t}=\text{161.7}\pm\text{3.0}$\,ps. Hereafter, the term time resolution refers to the resolution of the averaged signal.

Time resolution is limited by time walk and time jitter. Time walk is an effect caused by signal amplitude variations and the chosen triggering method for timestamping. Time jitter, on the other hand, depends on amplitude and noise fluctuations in the scintillator, photodetector, and readout system\,\cite{Leo1994}. Time jitter is expressed as $\sigma_{noise}/|dV/dt|$, where $\sigma_{noise}$ is the variance of the voltage signal due to noise and statistical fluctuations. This expression is approximated in terms of the rise time as $\sigma_{noise} t_{rise}/V_{signal}$. The effect of time walk on time resolution was minimized by the CDF technique applied to signal analysis as described in Sec.\,\ref{subsec:wave}. The effect of time jitter on the other side had to be examined at different overvoltages and temperatures.

\subsubsection{Temperature dependence}
\label{subsubsec:temp}

\begin{figure*}[t]
\begin{center}
\begin{tabular}{ll}
\includegraphics[width=6.5cm]{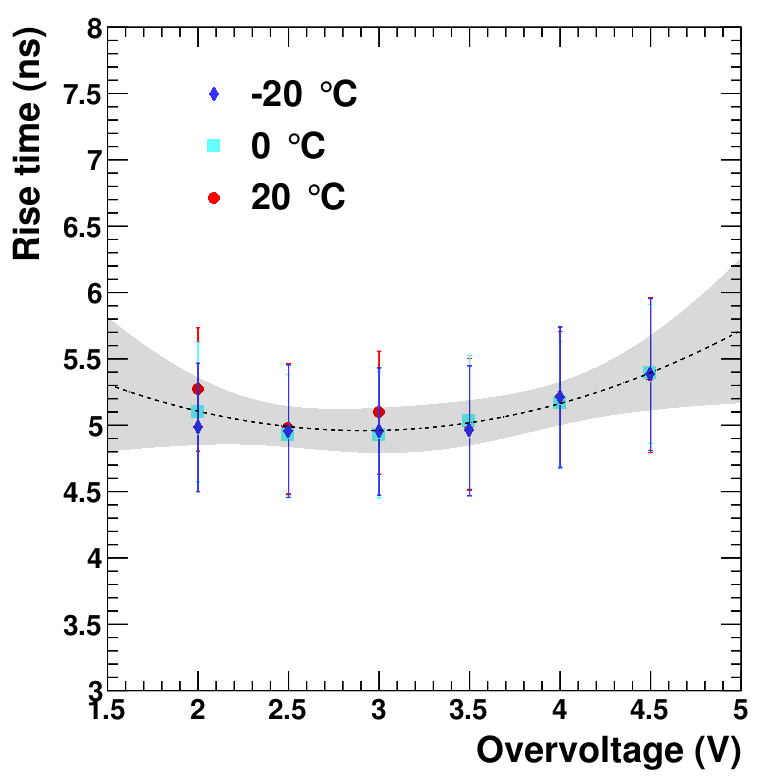}
& \includegraphics[width=6.5cm]{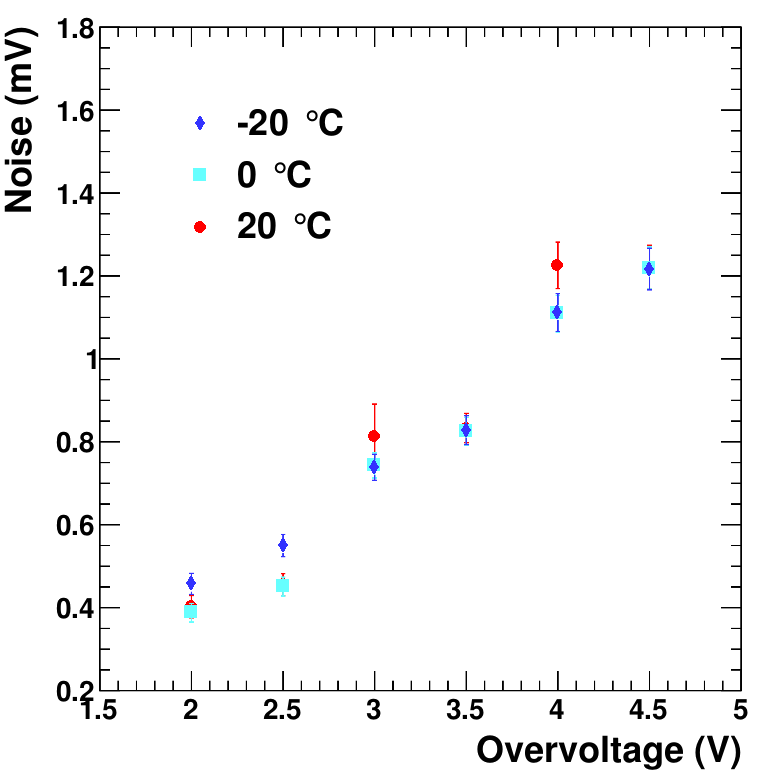}\\
\end{tabular}
\caption{Rise time (left) and noise signal (right) as a function of overvoltage for three representative temperatures.} 
\label{fig6}
\end{center}
\end{figure*}

Time resolution and, therefore, noise and rise time dependence on temperature were investigated using the experimental Setup\,1(see Sec.\,\ref{subsec:setup1}), where the SiPM array was operated under controlled thermal conditions. To further analyze time jitter, Figure\,\ref{fig6} shows the rise time (left) and noise (right) as a function of overvoltage for three specific temperatures. Note that the noise measured here was the baseline fluctuations, which already included both electronic noise (from amplifiers, cables, digitizer) and intrinsic SiPM noise (dark counts, correlated afterpulsing).

From the left panel of Figure\,\ref{fig6}, it can be seen that the rise time is minimum at a lower overvoltage, consistent with the faster development of the avalanche when the electric field is stronger. As overvoltage increases, particularly above 3\,V, secondary effects, such as abundant correlated avalanches and bandwidth limitations in the readout chain, can counteract this trend, resulting in the observed increase in rise time. A second--degree polynomial was fitted to the rise time data as shown by the dashed black line in Figure\,\ref{fig6} (left). The shadow area represents the 68\% confidence level of the fit. A minimum value of 5.0\,$\pm$\,0.2\,ns at $\text{V}_{\text{ov}}=\text{3\,V}$ was obtained from the polynomial fit. A systematic error of 1\% in the minimum value was estimated by testing the stability of the extracted points, varying the binning and fit range, propagating the resulting variations, and refitting.

On the other hand, from avalanche dynamics in SiPMs, it is expected that the noise increases with overvoltage\,\cite{Bonanno}, as observed on the right panel of Figure\,\ref{fig6}. This is a consequence of the higher avalanche triggering probability and the increasing gain of each avalanche with overvoltage, which enhances the signal amplitude but also amplifies fluctuations. This is observed in Figure\,\ref{fig6} (right), where noise rate increases with overvoltage for all temperatures. Although a dependence of the dark count rate on temperature is expected, these noise measurements are dominated by correlated noise (optical crosstalk and afterpulsing) and by the electronic readout floor, both of which depend primarily on overvoltage rather than temperature.

\begin{figure}[t]
\begin{center}
\includegraphics[width=7.0cm]{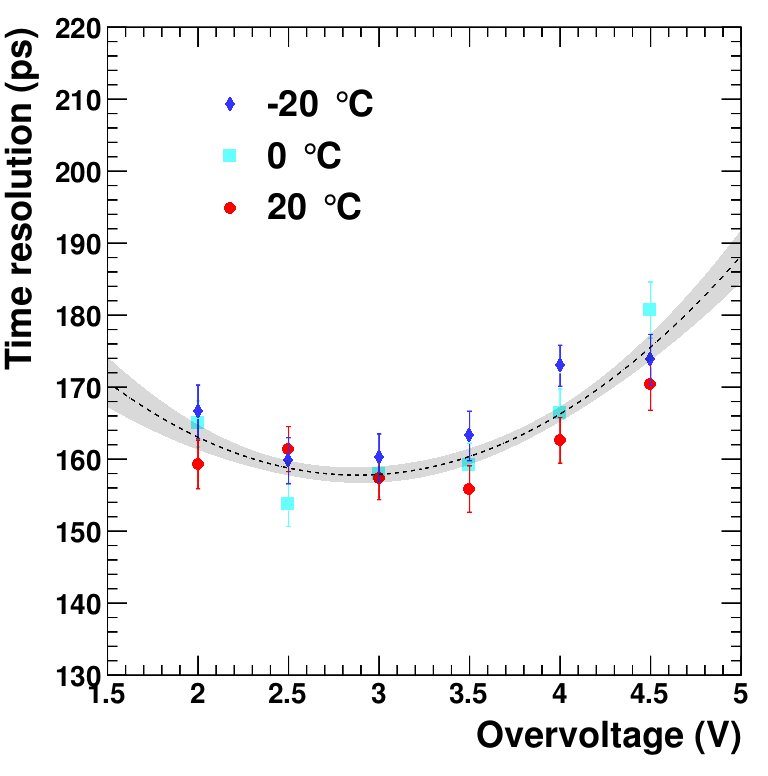}
\caption{Time resolution as a function of overvoltage for three representative temperatures.} 
\label{fig7}
\end{center}
\end{figure}

From the previous analysis, it can be anticipated that time jitter is minimum around 3\,V, hence, time resolution should be minimum near this overvoltage. This is observed in Figure\,\ref{fig7}, where the measured time resolution was plotted as a function of overvoltage for three different temperatures: $\text{-20}$\,°C, 0\,°C, and 20\,°C. Again, a second-degree polynomial was fitted to the data, obtaining a good $\chi^2/NDF=\text{1.36}$, with a minimum time resolution of 157.9\,$\pm$\,1.3\,ps at 3V.  A systematic error of 2\% in the minimum value was estimated by testing the fitting stability of time differences at the three temperatures under study. It is clear from Figure\,\ref{fig7} that the time resolution is minimum around 3\,V, and its behavior is similar across different temperatures, indicating that the breakdown voltage shift is correctly compensated.

\subsubsection{Position dependence}
\label{subsubsec:pos}

\begin{figure}[t]
\begin{center}
\includegraphics[width=8.0cm]{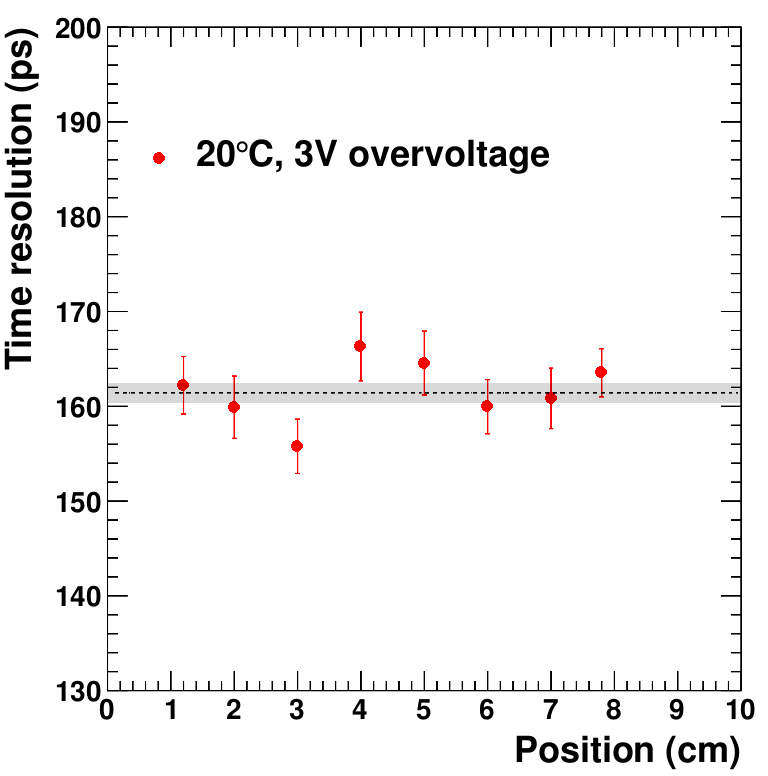}
\caption{Time resolution as a function of the source position.} 
\label{fig8}
\end{center}
\end{figure}

In addition to temperature dependence, time resolution was investigated as a function of the particle source's position along the plastic scintillation bar. For this study, the experimental Setup\,2 (see Sec.\,\ref{subsec:setup2}) was used, in which two SiPM arrays were coupled to opposite sides of P1 at a room temperature of 20\,°C. The $^{90}\text{Sr}$ source was placed every 1\,cm starting from 1\,cm on the left side of P1, maintaining a fixed overvoltage of 3\,V, identified as the optimal operating overvoltage from Sec.\,\ref{subsubsec:temp}. It was found that the time resolution varies slightly with respect to an average of 161.4\,$\pm$\,1.1\,ps, obtained from a linear fit ($\chi^2/NDF=\text{1.1}$) as shown in Figure\,\ref{fig8}, where the gray band represents the 68\% confidence interval. The systematic uncertainty from parameter variations in the time resolution fitting process was estimated to be 2\%. This result shows that, in this experiment, time resolution is not significantly affected by light attenuation or other propagation effects, as expected given the bar's short length.

\section{Conclusions}
\label{sec:conclusions}

In this work, a TOF prototype was developed using Onsemi MICROFC-30050-SMT SiPMs to study the dependence of time resolution on temperature and source position for applications in cosmic--ray instrumentation. Two experimental setups were used: the first to measure breakdown voltage and time resolution from $\text{-20}$\,°C to $\text{20}$\,°C using a cooling system based on a peltier module; and the second to measure the time resolution as a function of the source position at $\text{20}$\,°C. The temperature coefficient obtained from breakdown voltage vs temperature measurements was 22.0\,$\pm$\,1\,mV/°C, in agreement with values reported for similar SiPMs. A systematic error of 1\% was estimated by applying independent methods for breakdown calculation. The TOF prototype coupled to a 10\,cm BC418 plastic scintillator bar achieved time resolutions of 160-170\,ps over the temperature range, with optimal performance at 3\,V overvoltage and a minimum time resolution of $\sigma_{t}=\text{157.9}\pm\text{1.3}$\,ps. The resolution remained stable as the irradiation source was moved along the bar, demonstrating a uniform response across positions with an average value of 161.4\,$\pm$\,1.1\,ps. Fit--stability tests for rise time and time resolution introduced additional systematic uncertainties at the 1--2\% level. These results confirm the suitability of compact scintillator–-SiPM assemblies for precise timing measurements under varying thermal conditions, highlighting their potential for future balloon- and satellite--borne cosmic--ray experiments.

\section{Acknowledgments}
\label{sec:ack}
The authors thank the Electronics Department of the Instituto de Física, UNAM, in particular Maira Gloria Pérez Vielma, Rodrigo Alejandro Gutiérrez Arenas, and Jorge Israel Cruz Morales, as well as César Gustavo Ruiz Trejo and Eduardo López Pineda from the Medical Physics group. This work was supported by UNAM-PAPIIT IA101624 and by SECIHTI under grant CBF2023-2024-118.




 \bibliographystyle{elsarticle-num} 
 \bibliography{references}





\end{document}